\documentclass{aa}  

\usepackage{graphicx}
\usepackage{txfonts}
\usepackage{color}
\begin{document}

   \title{Modelling circumbinary protoplanetary disks}

   \subtitle{II. Gas disk feedback on planetesimal dynamical and collisional evolution in the circumbinary systems Kepler-16 and 34.}

   \author{S. Lines
          \inst{1}
          \and
          Z. M. Leinhardt\inst{1}\fnmsep
          \and
          C. Baruteau\inst{2,3}\fnmsep
          \and
          S.-J. Paardekooper\inst{4,5}\fnmsep
          \and
          P. J. Carter\inst{1}\fnmsep
          }

   \institute{School of Physics, University of Bristol, H. H. Wills Physics Laboratory, Tyndall Avenue, Bristol, BS8 1TL, UK\\
              \email{stefan.lines@bristol.ac.uk}
         \and
             CNRS, IRAP, 14 avenue Edouard Belin, 31400 Toulouse, France
          \and
             Universit{\'e} de Toulouse, UPS-OMP, IRAP, Toulouse, France
          \and
             Astronomy Unit, School of Physics $\&$ Astronomy, Queen Mary University of London, UK
          \and
             DAMTP, University of Cambridge, Wilberforce Road, Cambridge CB3 0WA, UK
             }

   \date{Accepted April 1st, 2016}

 
  \abstract
   {}
   {We investigate the feasibility of planetesimal growth in circumbinary protoplanetary disks around the observed systems Kepler-16 and Kepler-34 under the gravitational influence of a precessing eccentric gas disk.}
   {We embed the results of our previous hydrodynamical simulations of protoplanetary disks around binaries into an $N$-body code to perform 3D, high-resolution, inter-particle gravity-enabled simulations of planetesimal growth and dynamics that include the gravitational force imparted by the gas.}
   {Including the full, precessing asymmetric gas disk generates high eccentricity orbits for planetesimals orbiting at the edge of the circumbinary cavity, where the gas surface density and eccentricity have their largest values. The gas disk is able to efficiently align planetesimal pericenters in some regions leading to phased, non-interacting orbits. Outside of these areas eccentric planetesimal orbits become misaligned and overlap leading to crossing orbits and high relative velocities during planetesimal collisions. This can lead to an increase in the number of erosive collisions that far outweighs the number of collisions that result in growth. Gravitational focusing from the static axisymmetric gas disk is weak and does not significantly alter collision outcomes from the gas free case.}
   {Due to asymmetries in the gas disk, planetesimals are strongly perturbed onto highly eccentric orbits. Where planetesimals orbits are not well aligned, orbit crossings lead to an increase in the number of erosive collisions. This makes it difficult for sustained planetesimal accretion to occur at the location of Kepler-16b and Kepler-34b and we therefore rule out in-situ growth. This adds further support to our initial suggestions that most circumbinary planets should form further out in the disk 
   and migrate inwards.}


   \maketitle
%

\section{Introduction}

Circumbinary planets that orbit close to their parent stars form an interesting subset of the extreme planetary systems discovered to date. The strong gravitational perturbations that the stars exert on the protoplanetary disk can significantly alter the dynamics of both planetesimals \citep{lines14,paardekooper12,meschiari12a,thebault06} and gas \citep{lines15,paardekooper08,marzari13,pelupessy13} that orbit in close proximity to the system barycenter. Studies of the dynamics and collisional evolution of planetesimals in these hostile disks show that their eccentricities are pumped-up by these perturbations and their orbits misaligned, causing high velocity collisions during orbit crossing events \citep{lines14,paardekooper12,moriwaki04}. \cite{lines14} found that such a large number of high velocity encounters produces an overwhelming proportion of erosive collisions and showed that sustainable planetesimal growth was not possible at the orbital radius of the majority of discovered circumbinary planets.

The situation is more complex than these purely $N$-body studies reveal however, since planetesimals will feel both aerodynamic drag from, and the gravitational potential of, the gaseous component of the protoplanetary disk. A number of studies on the dynamics of the circumbinary gas disks show that the fluid is also perturbed by the binary, leading to the tidal truncation of the inner disk and the generation of spiral density waves by both direct forcing from the binary on the disk and mode coupling between the binary and disk potentials \citep{pierens13,lubow91}. This results in an eccentric precessing disk \citep{pierens13}. \cite{lines15}, which we will now refer to as Paper 1, explored a variety of fluid parameters for two known circumbinary planet systems, Kepler-16 and Kepler-34, and found that in all cases the inner disk became largely asymmetric with a build-up of gas at the disk apoapsis. Such regions of high surface density could present strong time-dependent gravitational forcing and drag on the planetesimals.

\begin{table}[t]
\centering
\vspace{+15pt}
\begin{tabular}{|c|c|c|c|c|c|c|c}
\hline 
\bf {Simulation} & \bf{System} & \bf{Distribution} & \bf{Gas Type} \\ \hline
\bf{A} & Kepler-16 & Unimodal & None\\
\bf{B} & Kepler-16 & Unimodal & Symmetric\\
\bf{C} & Kepler-16 & Unimodal & Asymmetric\\ \hline
\bf{D} & Kepler-16 & Mixed & None\\
\bf{E} & Kepler-16 & Mixed & Symmetric\\
\bf{F} & Kepler-16 & Mixed & Asymmetric\\ \hline
\bf{G} & Kepler-34 & Unimodal & None\\
\bf{H} & Kepler-34 & Unimodal & Symmetric\\
\bf{I} & Kepler-34 & Unimodal & Asymmetric\\ \hline
\bf{J} & Kepler-34 & Mixed & None\\
\bf{K} & Kepler-34 & Mixed & Symmetric\\
\bf{L} & Kepler-34 & Mixed & Asymmetric\\ \hline
\end{tabular}
\label{tab:gasmodels}
\vspace{+10pt}
\caption{Parameter setup of each simulation A through L. A-F are simulating a planetesimal disk around Kepler-16 and G-L around Kepler-34. Each of these eight simulations for each binary system are split into further subsets that consider a planetesimal unimodal and mixed size distribution.}
\end{table}

Due to the nature of performing simulations that unify both $N$-body and hydrodynamical effects, there is currently limited research on how the presence of a gas disk affects the dynamics of planetesimals and ultimately the collisions they undergo. \cite{rafikov13b} investigated how the gravitational potential from an axisymmetric disk affects embedded planetesimals, but we now know the disk is not in a static axisymmetric configuration. Additionally \cite{marzari13} examine both the gravitational and drag effects from a non-axisymmetric gas disk by running hybrid simulations but, they do not consider the interaction between the planetesimals themselves. Their work shows that planetesimal eccentricities are elevated by interaction with the gas disk, in some areas of the disk by up to ten times the value seen in simulations performed without gas disk gravity enabled.

In this paper we embed the hydrodynamic results of our Paper 1 into an $N$-body code to perform 3D, high-resolution, inter-particle gravity-enabled simulations of planetesimal dynamics that include the gravitational forces imparted by the gas disk. In section \ref{sec:method} we discuss our numerical method and initial conditions, in section \ref{sec:results} we present the results of our simulations and in section \ref{sec:dis} we discuss our results in the context of previous work.  

\section{Numerical methods}\label{sec:method}

Our $N$-body simulations are conducted using the widely adopted code PKDGRAV \citep{stadel01,richardson00}. PKDGRAV is a highly parallelised, multi-disciplinary code capable of handling large $N$. An efficient tree code and multipole expansion allows for $N$log$N$ scaling and hence the ability to run self-gravitating simulations with $N$ $\ge$ $10^5$ in a practical timeframe. The analytical collision model EDACM \citep{leinhardt12} which has been previously used in simulations of planet formation \citep{leinhardt15,carter15} is used to determine the outcome regime of a collision. EDACM takes the collision velocity, impact parameter, collider mass ratio and material parameters to assign the outcome as either perfect merging, partial accretion, non-erosive/erosive hit and run, erosive disruption and supercatastrophic erosion.

\subsection{Gas Potential}

To avoid the computational constraints of using a fully hybridised code, we take a semi-analytical approach to integrating the gas disk into the $N$-body code. The surface density data from 2D hydrodynamical simulations of circumbinary gas disks around Kepler-16 and Kepler-34 performed in Paper 1 are used.

\begin{figure}[t]
\centering
\includegraphics[scale=0.3]{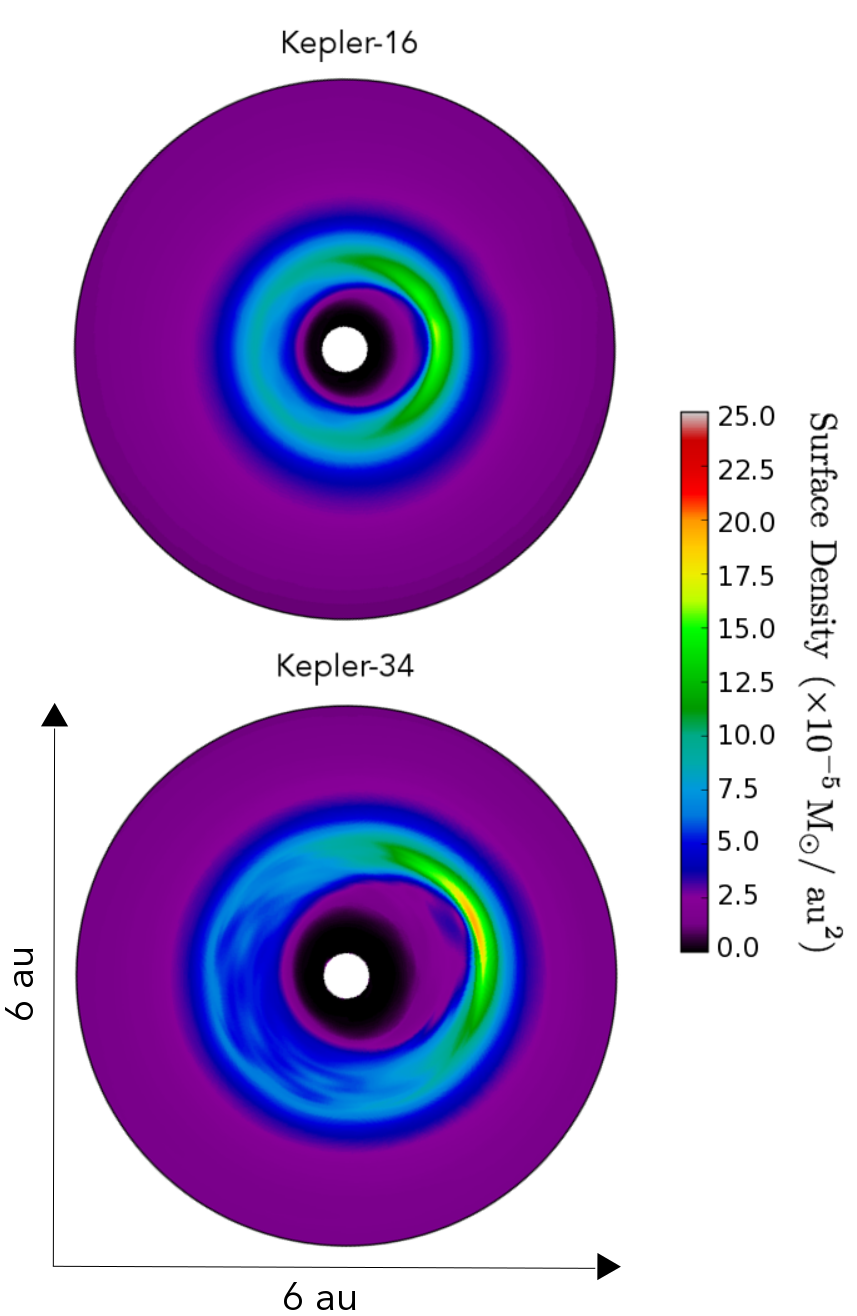}
\caption{Synthetic surface density maps of Kepler-16 and Kepler-34 created using the method described in Section \ref{sec:method}. The orientation of the disk is chosen to match its phase at the end of the $N$-body simulations in Figure \ref{fig:qss}.}
\label{fig:fluid}
\end{figure}

The quasi-steady-state (QSS) gas surface density data from Lines 2015 (simulations runs {\bf{B}} and {\bf{F}}) are used. The simulations correspond to non-self-gravitating disks since, for the disk masses explored here, the difference in gas surface density between self- and non-self-gravitating disks is marginal. These runs correspond to a locally isothermal disk with uniform aspect ratio, $h$ = 0.05, $\alpha$-viscosity, $\alpha$ = $10^{-3}$, initial surface density, $\Sigma(r) = 10^{-4} \left(\frac{r}{1 \textrm{au}}\right)^{-3/2} \textrm{M}_{\odot}/\textrm{au}^2$ (approximately half minimum mass solar nebula), a rigid inner boundary and no disk self-gravity. These fluid parameters are typical of circumbinary disk simulations. Self-gravitating gas disk simulations are ignored, since \cite{lines15} found that the gas disk only gravitationally interacts with itself for $\Sigma_{disk}$ $\geq$ 2.5 $\times$ $10^{-3}$ M$_{\odot}$/au$^2$. These hydrodynamical simulations are performed using FARGO-ADSG \citep{baruteau08,baruteau08t} over a polar mesh grid with $N_r$ = 395 and $N_s$ = 512 where $N_r$ and $N_s$ are the number of radial and azimuthal cells respectively. In Kepler-16 the gas disk has precession period of $P_d$ = 2000 P$_{AB}$ and in Kepler-34 has $P_d$ = 3000 P$_{AB}$ yielding $P_d \approx 250$ yr for both cases. The surface density maps of these disks can be seen in Figure \ref{fig:fluid} and are incorporated in the following way:

\begin{figure*}[t]
\centering
\includegraphics[scale=0.3]{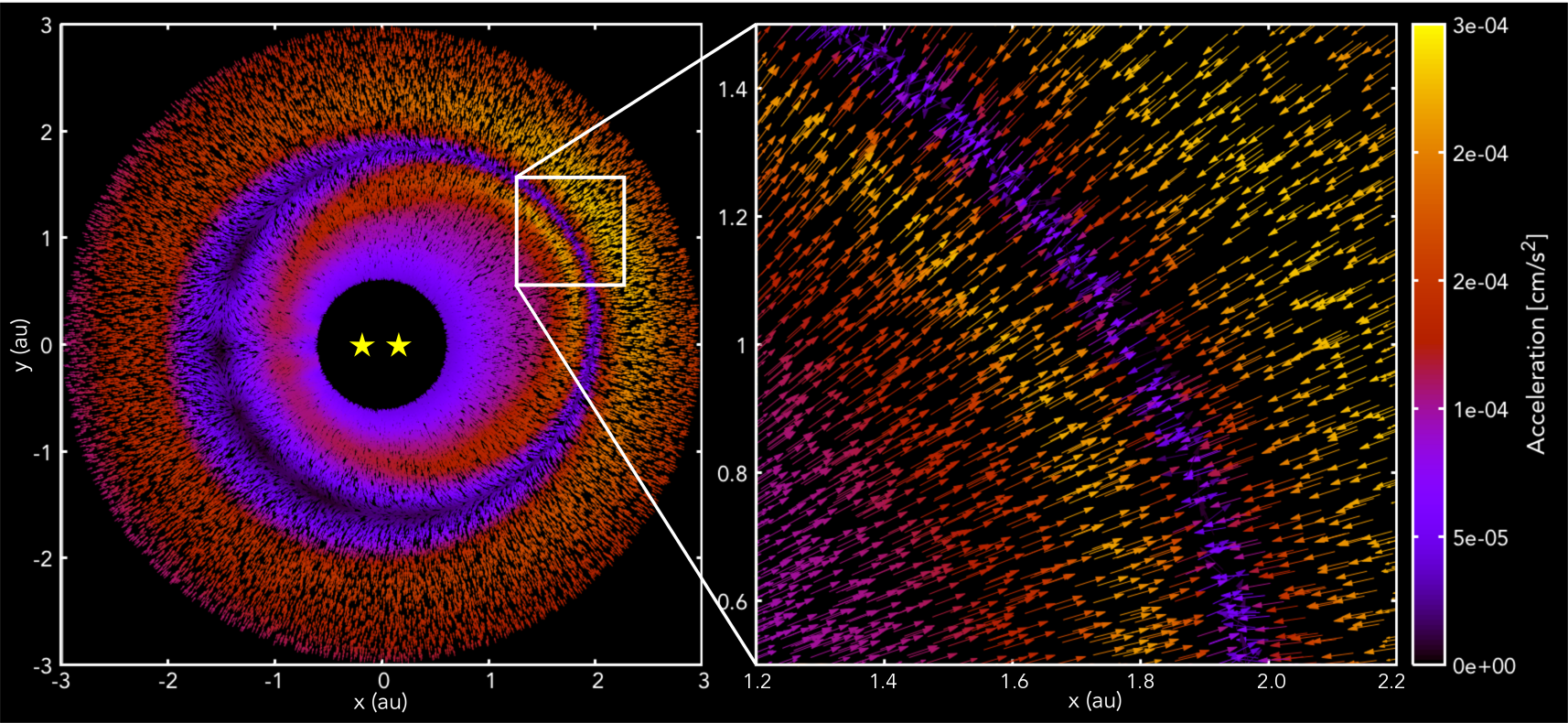}
\caption{Map of planetesimal accelerations from an $N$-body test simulation calculated from the gas disk potential of Kepler-34. 50$\%$ of planetesimals in the simulation are shown and are plotted as unit vectors with the colour corresponding to their acceleration magnitude.}
\label{fig:vec}
\end{figure*}
	
\begin{enumerate}  

        \item The disk surface density is averaged over one binary orbit to remove transient features such as short period modes. This is a valid approximation as a) the gas surface density evolves on a much longer timescale than a binary orbital period and b) short period modes are most active at the gas disk inner edge which is 0.3 au interior to the planetesimal inner edge and therefore would not have a significant impact of planetesimal dynamics.\\

        \item The gravitational potential for each cell $\Phi_{c}$ is calculated from the averaged cell surface density $\Sigma_{c}$ using direct summation. We use the Plummer potential $\Phi_c \propto (r^2+\epsilon^2)^{-1/2}$, with a smoothing length $\epsilon$ = 1.2$H$, where $H$ is the disk thickness, to smooth out singularities and to account for the vertical extent of the 2D disk \citep{muller12}.\\

        \item The gravitational potential is then Fourier transformed in angle into the lowest ten azimuthal modes (enough to generate a synthetic potential map that accurately describes the real output):
        
        \begin{equation}
        \Phi_m(r) = \int_{0}^{2\pi} \Phi_c(r,\phi)\textrm{exp}(\textrm{i}m\phi)\textrm{d}\phi 
        \end{equation}
where $\phi$ is the cell position angle and $\Phi_m$ is the complex form of the potential contribution, at a given radius, from each azimuthal mode $m$. At each radius the complex potentials are recorded in a table to be read into PKDGRAV.\\
        
\begin{figure*}
\centering
\includegraphics[scale=0.65]{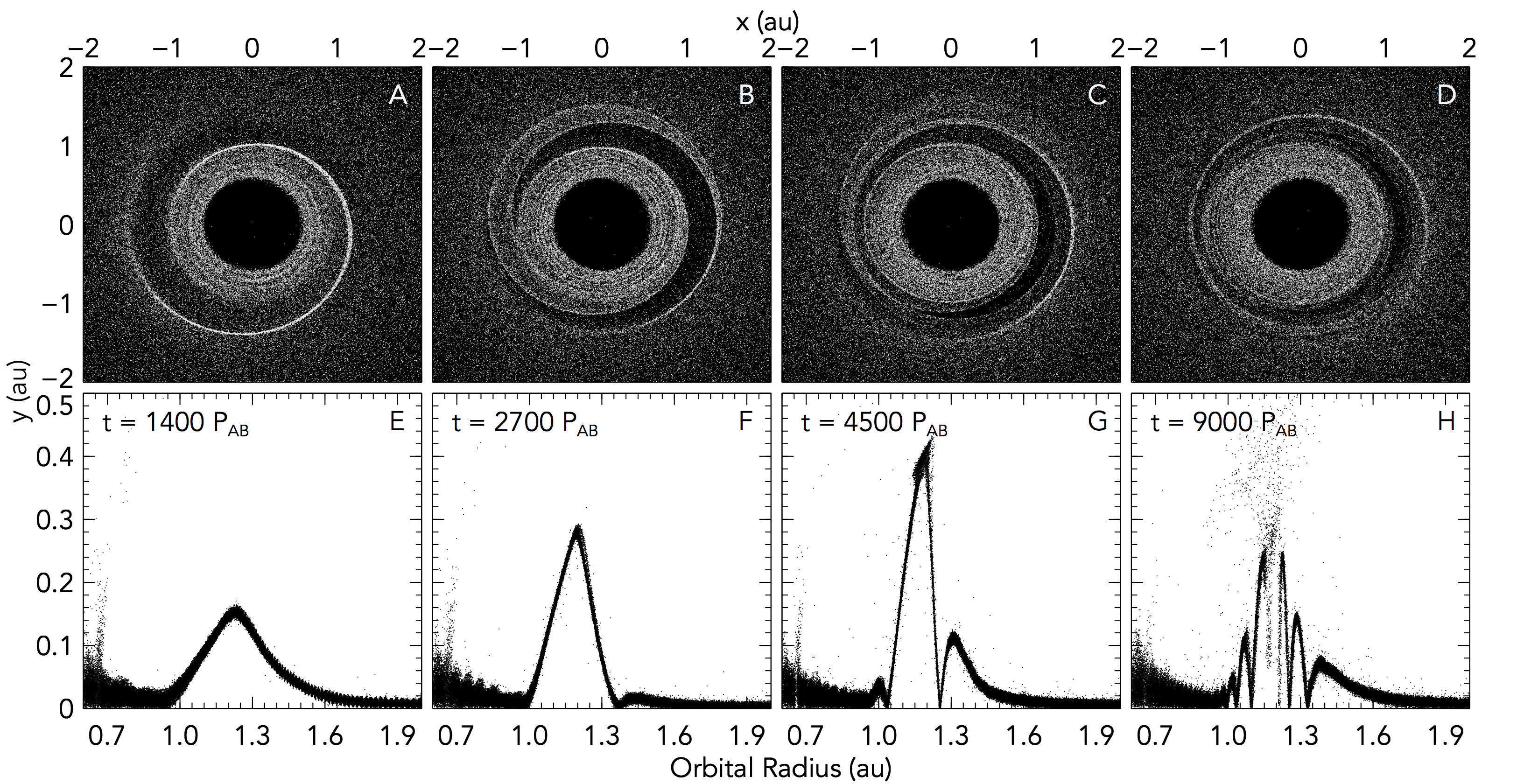}
\caption{Time evolution of planetesimal disk locations (top row) and eccentricity (bottom row) to quasi-steady-state around Kepler-16, under the influence of a full asymmetric gas potential.}
\label{fig:k16evo}
\end{figure*}
        
	\item PKDGRAV applies an acceleration on the planetesimals from the reconstructed gravitational potential of the gas disk. The gas disk potential must first be rebuilt by combining the complex modes. Mode combination is done as follows:
        
        \begin{equation}
        \begin{aligned}
\Phi(r,\phi,t) = \sum_{m = 0}^{m=10}&\Phi_m^r \textrm{Re}[\textrm{exp}(i(m\phi-\omega t))] - \\
      & \Phi_m^i \textrm{Im}[\textrm{exp}(i(m\phi-\omega t))] \\
\end{aligned}
\end{equation}

\vspace{+5pt}
where $\Phi_m^r$ and $\Phi_m^i$ are real (magnitude) and imaginary (phase) components of $\Phi_m$ respectively, $t$ is the simulation time and $\omega$ is the precession frequency. This method retrieves the disk gravitational potential for any value of $r$, $\phi$ and $t$.\\

\item The gas potential and acceleration are evaluated at the particles location using a bilinear interpolation.\\

\item This additional acceleration is added onto the existing acceleration of the planetesimal calculated in PKDGRAV's other functions which evaluate the force on the planetesimals from inter-planetesimal and stellar-planetesimal interactions. The result of this implementation is shown in Figure \ref{fig:vec} where the planetesimal accelerations in response to the potential of a Kepler-34 gas disk are shown.\\
\end{enumerate}
For simulations where we test the static, symmetric circular disk only ($m$ = 0), the same procedure applies but values of the azimuthal mode number greater than zero are ignored. The suite of simulations defined in Table \ref{tab:gasmodels}, explores the effects of three different gas profile types and varying the size distribution of the planetesimal disk.

\begin{figure*}
\centering
\includegraphics[scale=1.4]{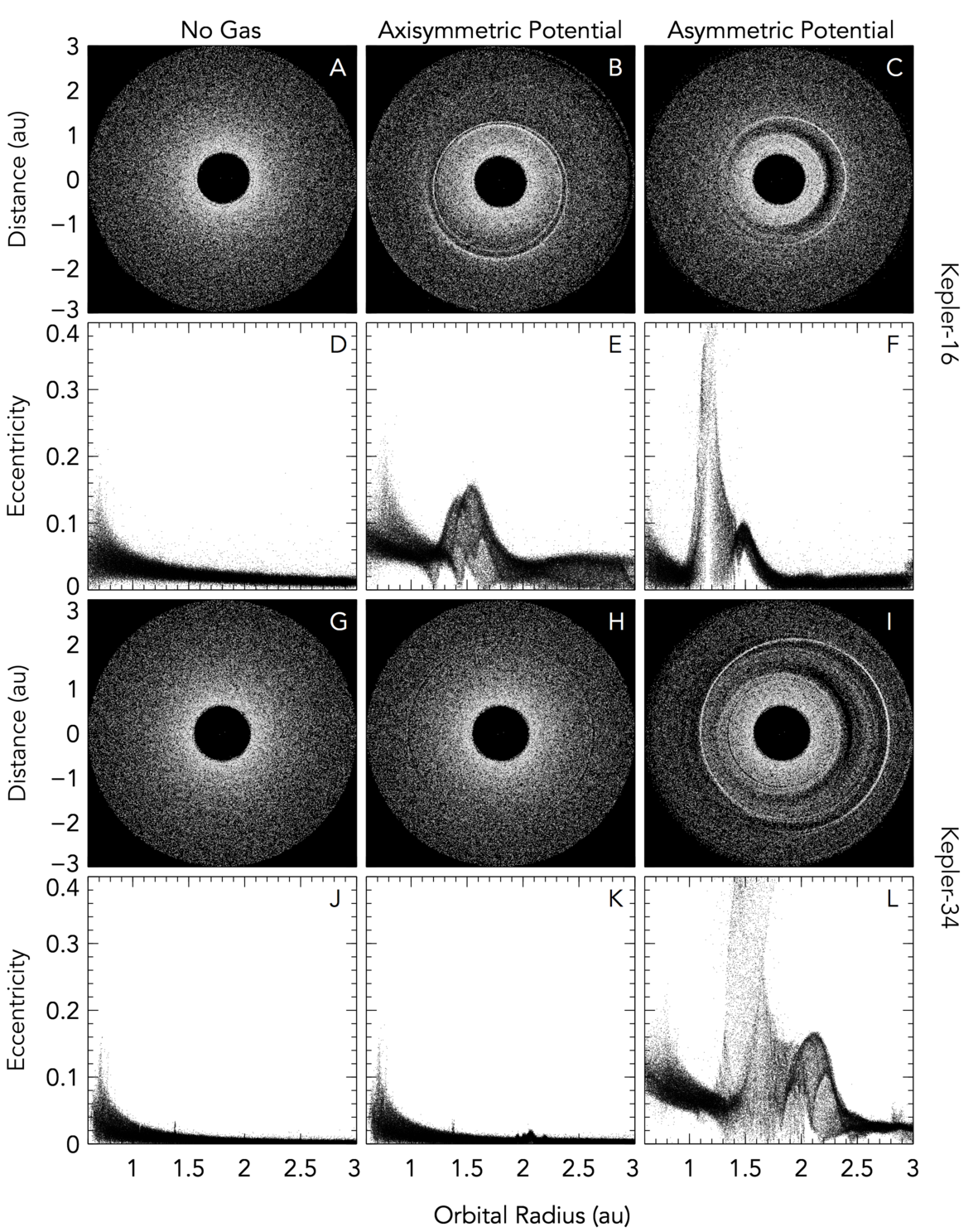}
\caption{Planetesimal disks at quasi-steady-state (end of simulation at t $\approx$ 40,000 $P_{AB}$. A-F: Location and eccentricity of planetesimals in disk around Kepler-16. G-L: Location and eccentricity of planetesimals in disk around Kepler-34.}
\label{fig:qss}
\end{figure*}

\begin{figure*}[t]
\centering
\includegraphics[scale=0.7]{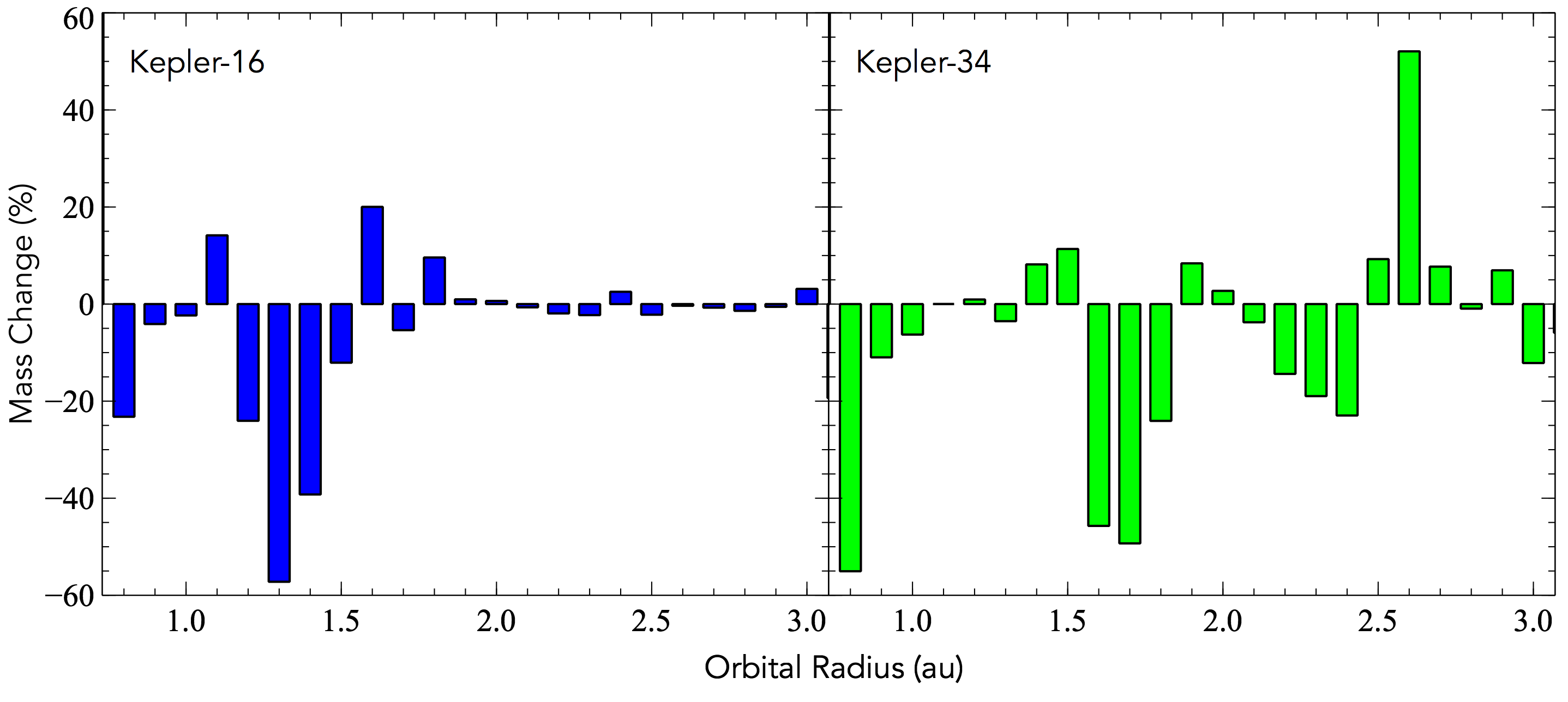}
\caption{Change in the distribution of planetesimal mass between initial conditions and simulation end for the asymmetric potential only cases. Planetesimals are placed into bins of width 0.1 au and the increase or decrease of mass in each bin is plotted as a percentage change.}
\label{fig:mass}
\end{figure*}

\subsection{Initial Conditions}

We test the response of a circumbinary planetesimal disk around two different stellar binary systems, Kepler-16 \citep{doyle11} and Kepler-34 \citep{welsh12} which have a total mass of 0.89 $M_\odot$ and 2.07 $M_\odot$ respectively. Kepler-34 has $M_{\textrm{A}}$ = 1.05 $M_\odot$ and $M_{\textrm{B}}$ = 1.02 $M_\odot$ where $M_{\textrm{A}}$ and $M_{\textrm{B}}$ are the primary and secondary stellar mass, stellar separation is $a_{\textrm{b}}$ = 0.22 au and binary eccentricity is $e_{\textrm{b}}$ = 0.53. Kepler-16 has $M_{\textrm{A}}$ = 0.69 $M_\odot$ and $M_{\textrm{B}}$ = 0.20 $M_\odot$, $a_b$ = 0.22 au and $e_b$ = 0.16. Each planetesimal disk has a total mass of 2.8 $M_\oplus$ spread over $N$ = $10^5$ planetesimals and the disk domain ranges from $R_{\textrm{in}}$ = 0.6 to $R_{\textrm{out}}$ = 3.0 au. The planetesimal density is 2.0 g/cm$^{3}$ to match that of silicon-rich rocky bodies. Two different models are used for the planetesimal disk size distribution: unimodal and mixed. For the unimodal size distribution, $m_{\textrm{p}}$ = 8.3 $\times$ 10$^{-11}$ $M_\odot$. For the mixed size distribution planetesimals can occupy one of ten mass bins that are equal in their total mass and range from $m_{\textrm{p}}$ = 3.8 $\times$ 10$^{-11}$ - 9.7 $\times$ 10$^{-9}$ $M_\odot$. The large size of these planetesimals is a necessary condition; simulating much smaller bodies would require either the impractical computational demand of increasing the $N$-body resolution or reducing $M_{\textrm{disk}}$ for constant $N$ and thus reducing the collision rate. We want to ensure realistic collision timescales so that any net accretion rate can be compared to the overall lifetime of the protoplanetary disk. Additionally our planetesimal masses are an order of magnitude larger than in \cite{lines14}, as a decrease in the $N$-body resolution for this work is necessary to evolve the disk to a quasi-steady-state in a practical timeframe. Such large planetesimals, which may be able to form at this size through fast clumping assisted by streaming-instabilities \citep{johansen07,carrera15}, provide a convenient best-case scenario as they have a large gravitational binding energy that makes them difficult to disrupt. Should it be found that even they are not able to undergo accretion, smaller bodies would only disrupt more easily. The planetesimal surface density follows $\Sigma(r) \propto r^{-1.5}$ which is consistent with our previous $N$-body study of circumbinary planetesimal disks \citep{lines14}.

\begin{figure}[t]
\centering
\includegraphics[scale=0.65]{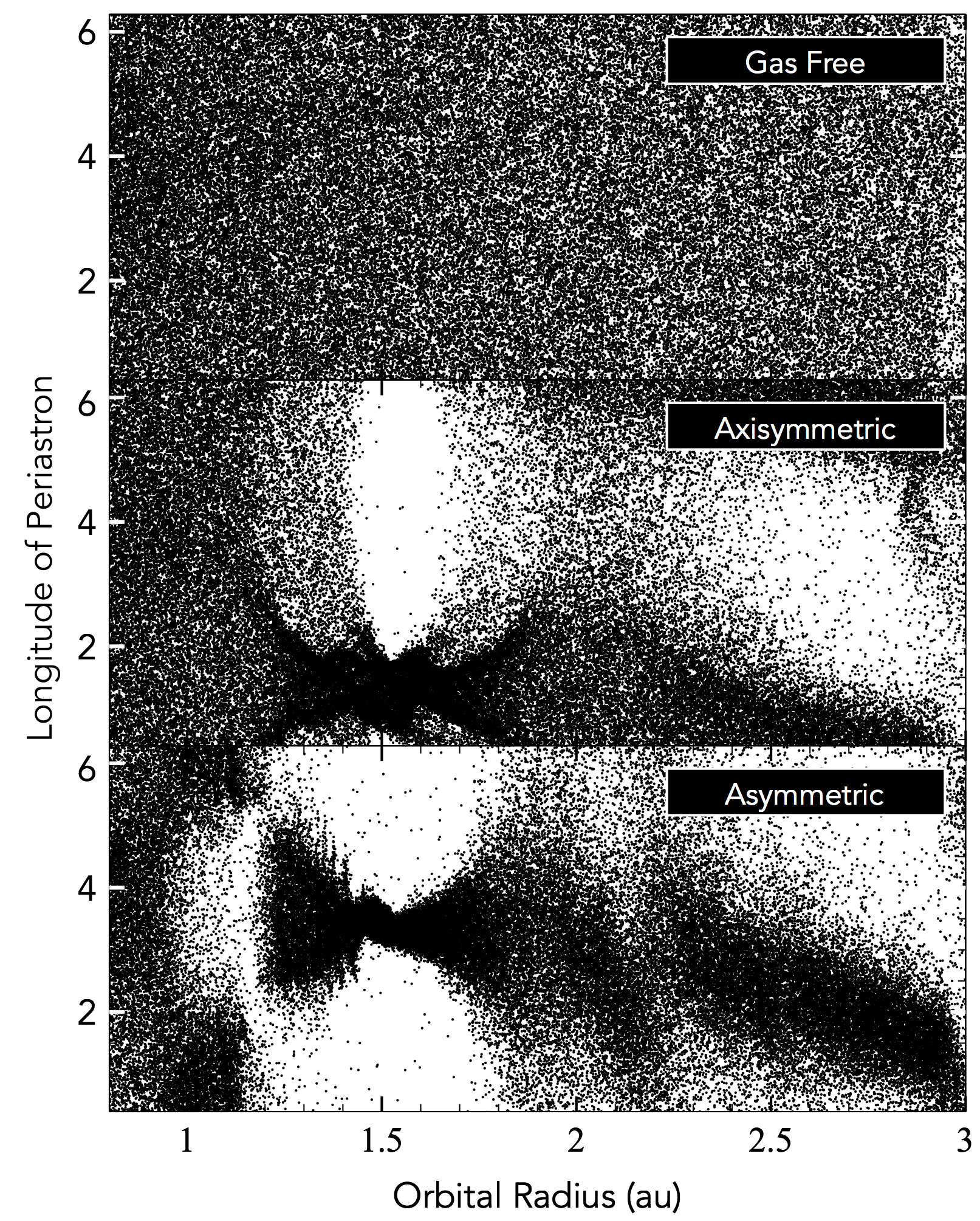}
\caption{Longitude of periastron for each planetesimal (black dots) in the Kepler-16 system.}
\label{fig:lop}
\vspace{-10pt}
\end{figure}

A small value for the timestep is required (0.0025 yr) to accurately resolve the binary, which is modelled as two interacting $N$-body particles, and maintain stability over thousands of binary orbits. As per \cite{lines14} we begin each simulation with an unperturbed planetesimal disk with the inclinations and eccentricities assigned from a Rayleigh distribution with dispersions of $\langle e\rangle^2 = 2\langle i\rangle^2 = 0.007$ \citep{ida92}. We enable collisions, and gas gravity if applicable, from t = 0 and allow the disk to reach a quasi-steady-state. Only collisions that occur from after a quasi steady-state are used in the analysis and results. Each simulation runs for 3800 years in total (34,000 $P_{AB}$ and 50,000 $P_{AB}$ for Kepler-16 and Kepler-34 respectively), a small snapshot of the million year timescale over which planet formation occurs.

\section{Results}\label{sec:results}

\begin{figure}
\centering
\includegraphics[scale=0.6]{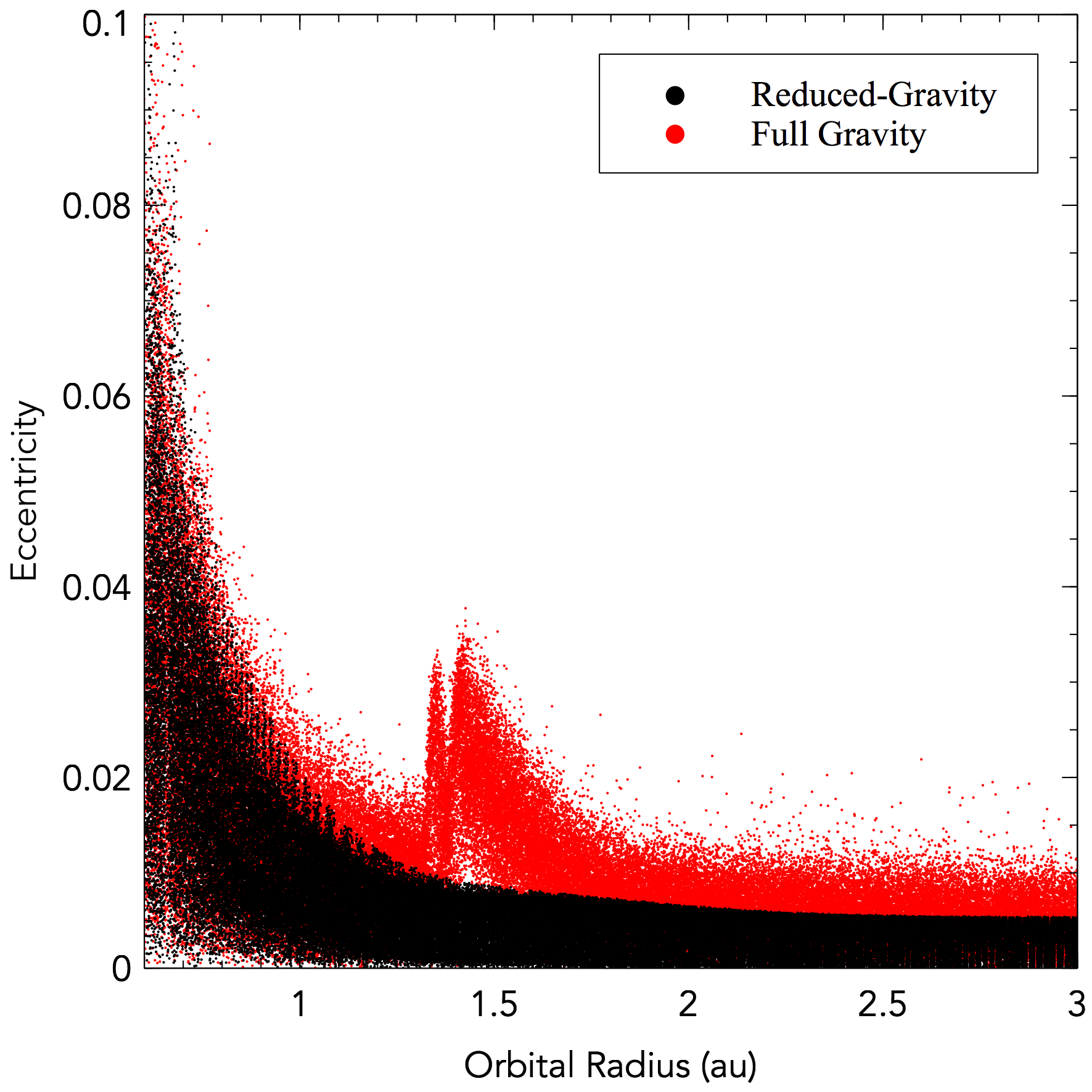}
\caption{Change in planetesimal eccentricities by reducing planetesimal self-gravity. Snapshot at 14,000 $P_{\textrm{AB}}$ for Kepler-16 with an axisymmetric gas disk..}
\label{fig:grav}
\end{figure}

\subsection{Gas Potential}

\subsubsection{Full asymmetric disk potential}

We start by looking at how the presence of the gas disk potential modifies the dynamics of the planetesimals. The strongest gravitational influence comes from areas of highest surface density, most significant around the gas disk inner edge apoapsis. At apoapsis of an eccentric orbit the gas has its lowest velocity which leads to an inevitable over-density of fluid. The small region over which this over density exists means that it acts almost like a massive body which stirs up planetesimal eccentricities as they traverse though it. Planetesimals that orbit exterior to the density peak are attracted towards it and move inwards while those that orbit interior to it move outwards. This can be seen in Figure \ref{fig:vec} by the strong acceleration directed towards the dense fluid peak (the peak itself contains planetesimals with very low accelerations due to being at the point of highest potential). This ultimately leads to the formation of a narrow annulus of planetesimals that can cause a separate interaction with surrounding planetesimals due to their self-gravitating nature. 

Figure \ref{fig:k16evo} shows the formation and evolution of this feature along with the associated planetesimal eccentricities, for an asymmetric gas potential on a Kepler-16 disk. In panel A the narrow annulus or arm takes on a spiral form due to the planetesimals experiencing a shift from positive to negative torque as they transit across the density peak. Since $\omega << \omega_p$ where $\omega_p$ is the planetesimal orbital frequency, planetesimals make two orders of magnitude more orbits in the time it takes for the eccentric pattern in the gas disk to make one full precession. The planetesimals are thus concentrated quickly with the transient spiral disappearing after only a few thousand binary orbits and before the gas disk has rotated twice.

In panel D at 9000 $P_{AB}$ the planetesimal disk has almost reached a quasi steady state with an eccentric planetesimal ring that passes through the density peak and more noticeably an asymmetric cavity interior to the ring almost devoid of planetesimals due to their migration outwards. Planetesimals interior to around 0.9 au do not strongly feel the gas gravity and lead relatively unperturbed orbits, with their eccentricities primarily set by the binary forcing (for a full description of planetesimal dynamics around binaries, see \cite{lines14}). The presence of the gas potential leads to eccentricity waves that are launched outside of the density peak region and move inwards, bunching up at the peak at 1.2 au. These eccentricity waves eventually compress together to form a smooth distribution of planetesimal eccentricities which can be seen at the end of the simulation in panel F of Figure \ref{fig:qss} at 40,000 $P_{AB}$. The planetesimal disk under the influence of the full asymmetric gas potential has a much higher mean disk eccentricity than the gas free case. Around the density peak, eccentricities reach a maximum of $e_{\textrm{max}}$ = 0.4. This is four times that found in the gas free simulation ($e_{\textrm{max}}$ = 0.1), which occurs at the innermost planetesimal disk edge and is set by the binary forcing. At the end of the simulation as can be seen in panel F of Figure \ref{fig:qss} planetesimal eccentricities are enhanced from 1.0 to 1.7 au.

\begin{figure*}[t]
\centering
\includegraphics[scale=0.65]{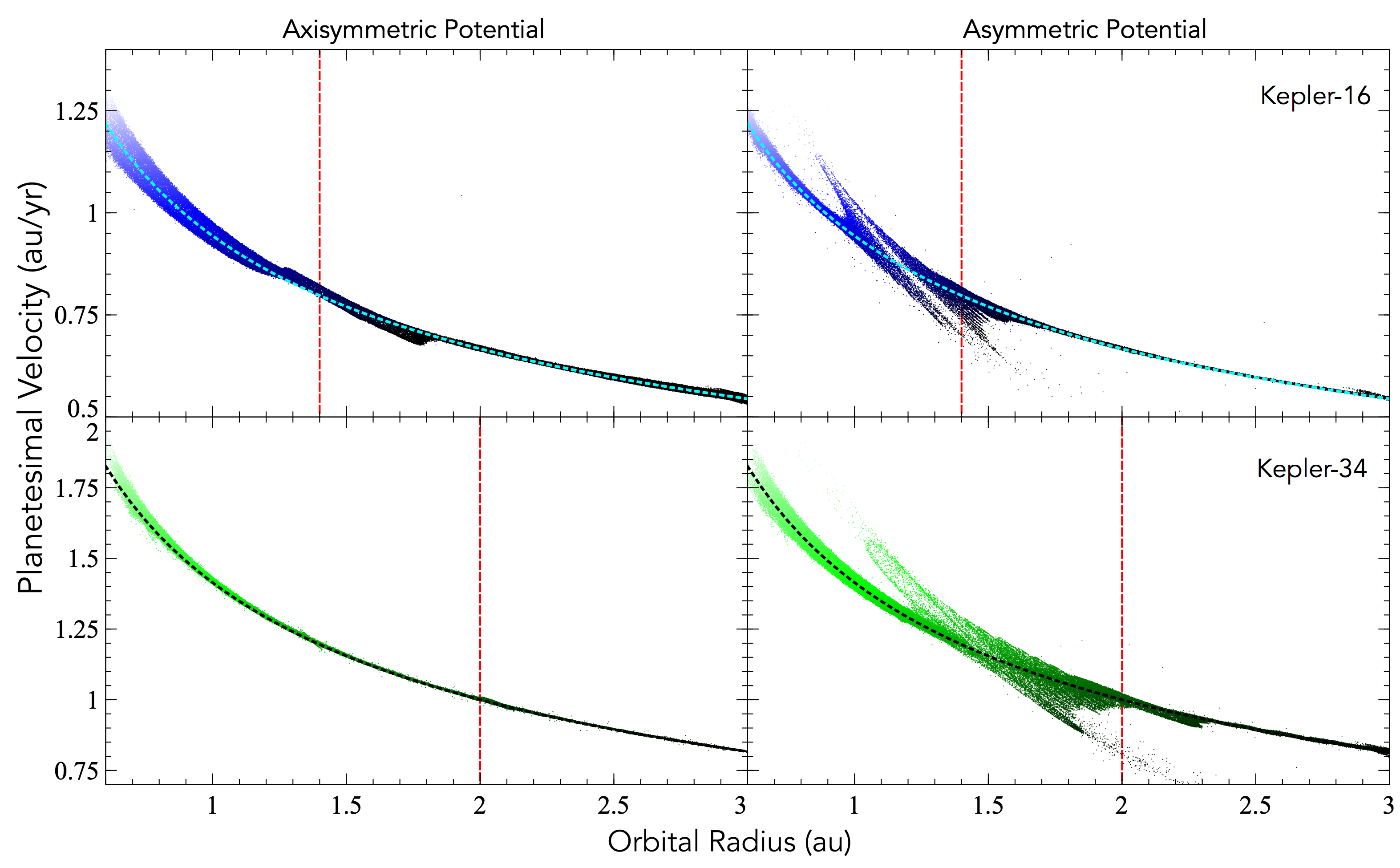}
\caption{Planetesimal orbital velocity in Kepler-16 (top row) and Kepler-34 (bottom row) for both axisymmetric and asymmetric gas disk gravity inclusion. The Keplerian velocity is shown as a dotted curve and the location of the gas surface density peak at apoapsis (over-density of fluid region) is shown as a red dashed line.}
\label{fig:orbvel}
\end{figure*}

Figure \ref{fig:mass} assists in understanding the change in planetesimal mass distribution under the influence of the gas disk gravity. In Kepler-16 the cavity is shown by the negative mass change between 1.1 and 1.5 au. The high density planetesimal ring appears as a small increase in planetesimal mass just exterior to this deleted region. In Kepler-34 there is a more defined planetesimal ring shown by the large mass increase at 2.6 au, corresponding to the fluid density peak. The cavity stretches from 1.6 to 2.5 au but is not fully depleted in the middle since planetesimals in this region has a low acceleration from the gas potential due to their distance from any significant fluid mass.

Alignment of planetesimal pericenters ($\varpi$) (see Figure \ref{fig:lop}) is another effect contributed by the presence of gas disk gravity. The gas free case shows no alignment at all resulting in overlapping eccentric orbits throughout the disk, but the eccentricity of the planetesimals themselves is low. The orientation of the orbits becomes much more important when the eccentricities are raised to those values seen in the presence of the full asymmetric disk gravity. The asymmetric cavity is represented by the lack of planetesimals on one side of the disk, with low occupancy between $\varpi$ $\approx$ 2 and $\varpi$ $\approx$ 5 in the region of $a$ = 1.0 - 1.2 au. The density peak that focuses planetesimals onto highly eccentric orbits also neatly aligns them with $\Delta \varpi \approx$ 0.6 between 1.4 au and 1.6 au. The gas fails to align planetesimals between 1.2 au and 1.4 au however and leads to a wide range of orbit orientations, $\Delta \varpi \approx$ 3.

Planetesimals in the region around and interior to the density peak experience a departure from their Keplerian velocity. It can be seen in Figure \ref{fig:orbvel} that the velocity distribution becomes winged for both Kepler-16 and Kepler-34. Near the peak some planetesimals obtain a much lower orbital velocity - these are the planetesimals captured onto an eccentric ring focused by the density peak, with low orbital velocities calculated at apoapsis. When these planetesimals transit through periapsis their velocities are increased, shown by the positive velocity wing.

\begin{figure*}[t]
\centering
\includegraphics[scale=0.75]{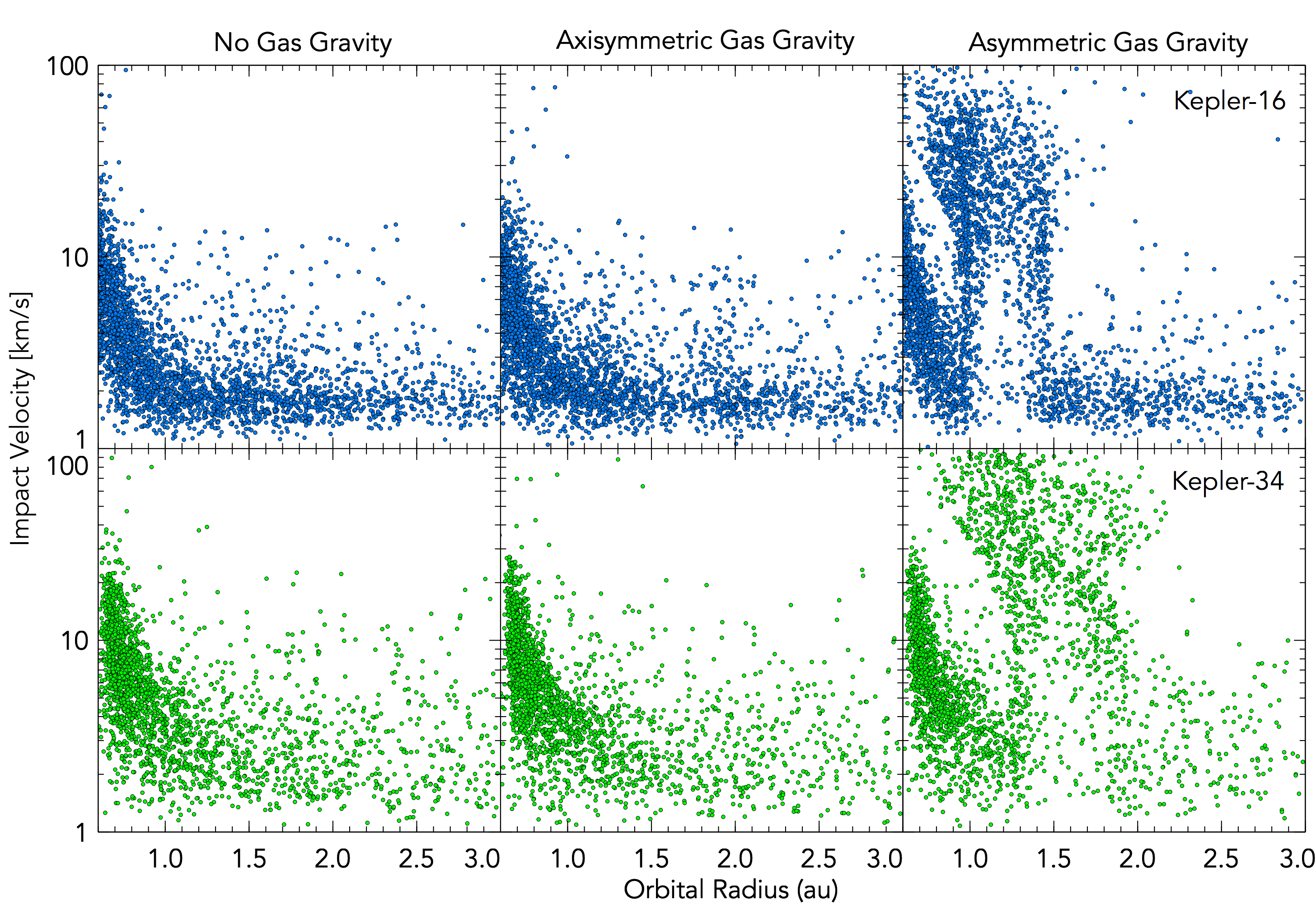}
\caption{Impact velocity for collisions in disk around Kepler-16 (top row) and Kepler-34 (bottom row).}
\label{fig:collvel}
\end{figure*}

A combination of eccentricity waves generated by the asymmetric potential and inflated orbit velocities leads to significantly higher impact velocities between planetesimals during collisions. In Figure \ref{fig:collvel} the impact velocities can be seen to exceed ten times the typical value seen for the gas free case. The impact velocity distributions, shown in Figure \ref{fig:collvel} are enhanced around the position of the orbital velocity wings since this is the location where planetesimals can obtain either the Keplerian value or the inflated or deflated value caused by the potential of the disk, and hence the relative velocities during collisions can be large.

\begin{figure*}[ht]
\centering
\includegraphics[scale=0.8]{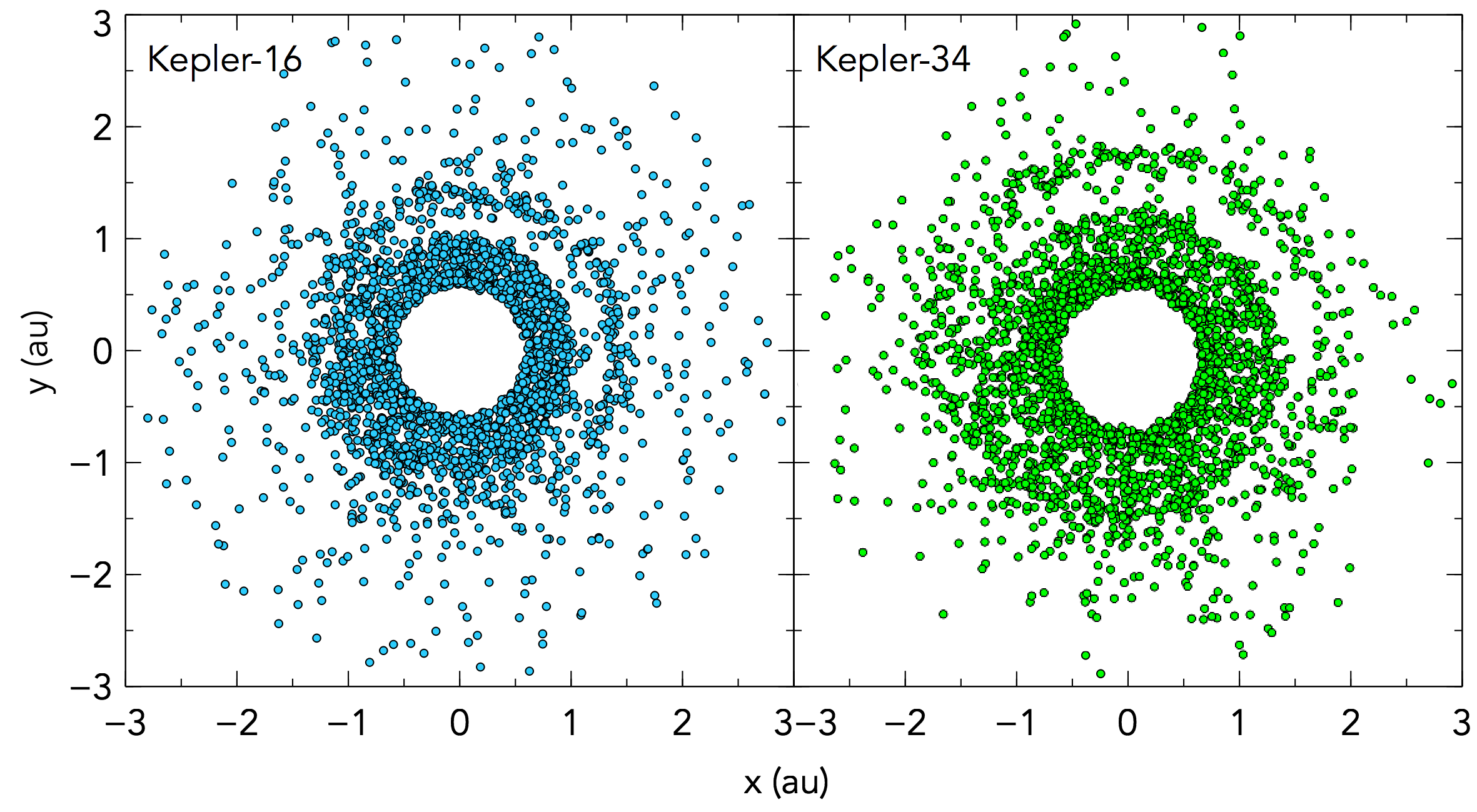}
\caption{Spatial distribution of collisions recorded in the simulation for the asymmetric gas potential case of both Kepler-16 (left) and Kepler-34 (right). The position of each collision is rotated by the gas disk phase to produce a relative position.}
\label{fig:collmap}
\end{figure*}

Spatial collision maps (Figure \ref{fig:collmap}) can be used to show that the asymmetric cavities chiseled out by the strong gravitational pull of the high density ring exterior to it, are as devoid of collisions as they are of planetesimals. Collision rates drop off with distance from the binary even without gas, due to the drop in planetesimal number density, so the addition of this no-collision zone increases the difficulty for planetesimals to grow.

The most damning data in terms of the ability for planetesimals to growth is the collision type occurrence. A collision is put into one of 24 bins with a width of 0.1 au, and at the end of the simulation the percentage of each collision type can be displayed as a function of orbital radius. This can be seen for Kepler-16 in Figure \ref{fig:k16coll} and for Kepler-34 in Figure \ref{fig:k34coll}. For Kepler-16 the presence of the full asymmetric gas gravity causes a significant enhancement in the number of erosive collisions from the disk inner edge at 0.6 au to 1.8 au. Considering the mixed size distribution at the location of the planet Kepler-16b at 0.7 au, 40$\%$ of all collisions are disruptive to one of the colliders, with this increasing to 70$\%$ between 1.0 au and 1.3 au. For Kepler-34b at 1.1 au, 30$\%$ of collisions are disruptive with this increasing to 75$\%$ between 1.4 au and 1.6 au. Only exterior to 2.1 au does the number of growth enabling collisions exceed that of destructive ones.

\subsubsection{Axisymmetric Gas Potential}

For the axisymmetric case, the disk is static in time with the gas surface density smeared out azimuthally. This means that the density peak in the full asymmetric description is averaged out over an annulus leading to a lower density ring. However the focusing effect of the ring on the planetesimals is comparatively low with respect to the asymmetric case since the high density has been averaged out. For Kepler-16, despite the dense gas ring having $e$ = 0.0, the gravitational forcing from the binary causes the planetesimals to form a static double ring with $e$ > 0.0 which is clearly seen in panel B of Figure \ref{fig:qss}. The eccentricity distribution in panel E shows that $e_{\textrm{max}}$ = 0.18 is much smaller than that found in the full asymmetric case.

An interesting observation is that in Kepler-16, planetesimal eccentricities obtain much higher values than the gas free case, and particularly for the region beyond where the gas potential is strong ($>$ 2 au). This is contradictory to the results of \cite{rafikov13b} who find that the gravitational contribution from an axisymmetric disk helps to reduce excited planetesimals. To try and understand this further we run an additional simulation, identical to that of Run B (Axisymmetric, Kepler-16) but reducing the mass of planetesimal by a factor of 10$^5$ (decreasing a planetesimals radius by a factor of $\approx$ 50). This reduces the gravitational interaction between them, essentially leaving us with the motion of test particles about the binary and subject to the potential of the gas. The results, shown in Figure \ref{fig:grav}, clearly show that the self-gravitation of our planetesimals has a large contribution to the dynamics. Not only does reducing inter-particle gravity remove the high eccentricities at the disk outer edge, but it also removes the peak associated with the high density gas disk ring.

In Kepler-34 seen in panel H of Figure \ref{fig:qss} the effects are far less noticeable since the gas ring is further out and hence less dense. This means the gravitational effect is weaker, but also acts on a smaller number of planetesimals since the number density of planetesimals falls off with distance from the binary. The eccentricity distribution barely diverges from the gas free case with $e_{\textrm{max}}$ being set by the binary forcing and not from the presence of gas. The ring is static and not eccentric, unlike that seen in Kepler-16.

\begin{figure*}[ht]
\centering
\includegraphics[scale=0.65]{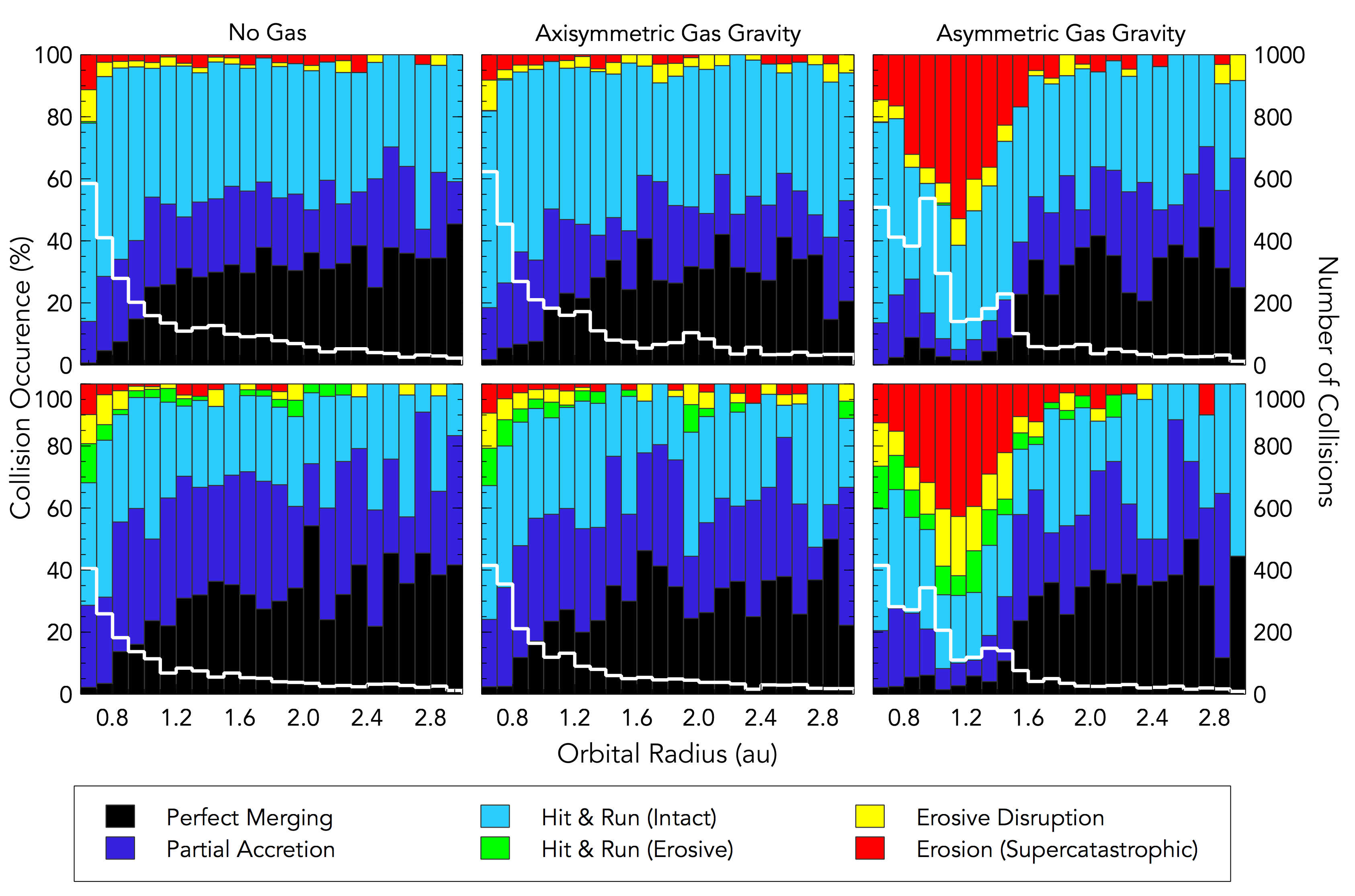}
\caption{Collision type occurrence for gas free, static axisymmetric gas potential and asymmetric gas potential simulations for a unimodal (upper) and mixed (lower) size distribution disk around Kepler-16. White line shows total number of collisions.}
\label{fig:k16coll}
\end{figure*}

For Kepler-16, the gas surface density maximum, shown in Figure \ref{fig:qss}, at 1.5 au is efficient at aligning planetesimal pericenters even in the axisymmetric case. As can be seen in Figure \ref{fig:lop}, alignment is at a maximum at 1.5 au but begins from 1.2 and continues to the outer edge of the planetesimal disk. The axisymmetric case sees a dephasing of orbits at around 2.0 au, similar to that seen when considering the full asymmetric gas potential.

\begin{figure*}[ht]
\centering
\includegraphics[scale=0.65]{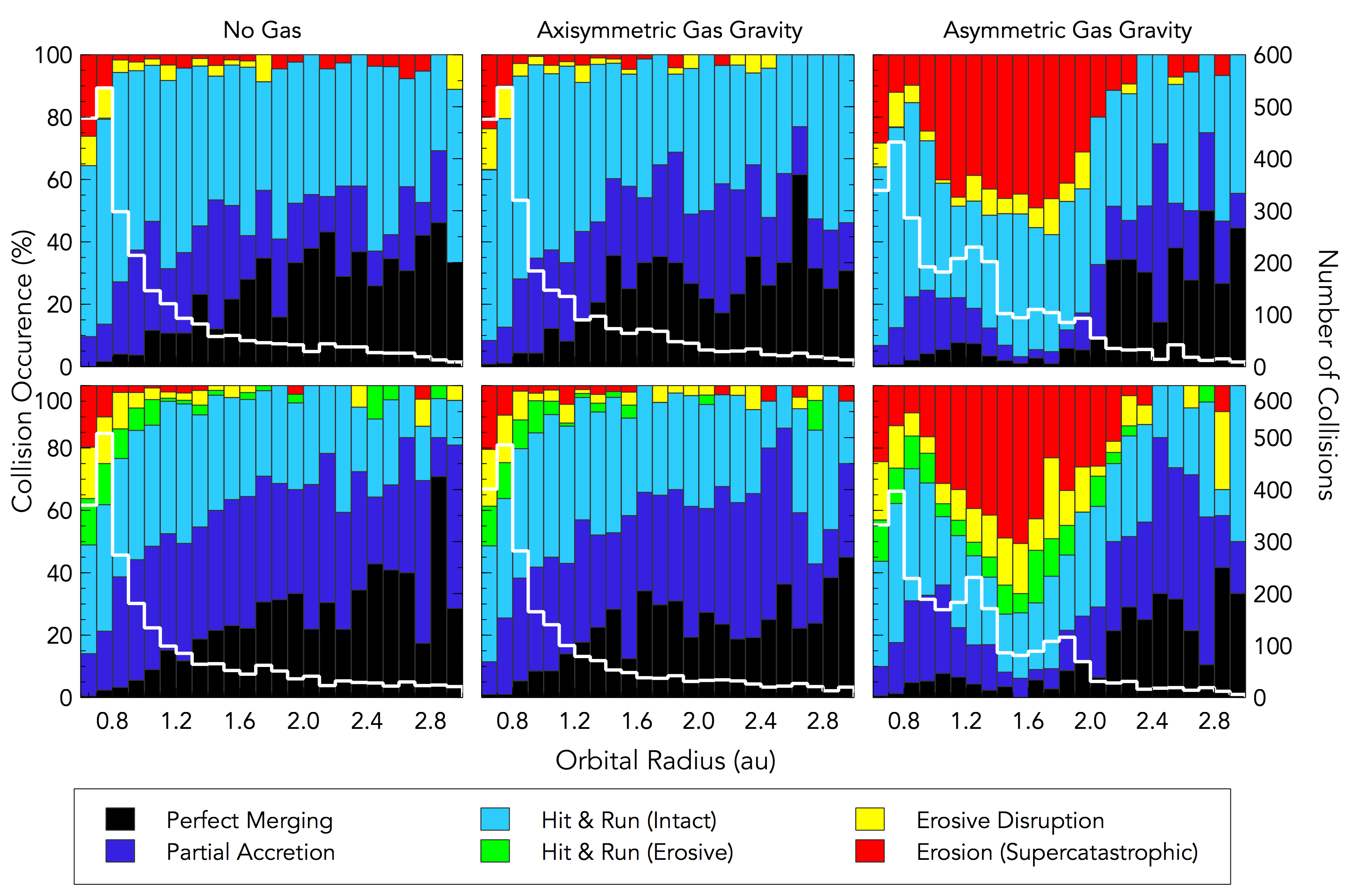}
\caption{Collision type occurrence for gas free, static axisymmetric gas potential and asymmetric gas potential simulations for a unimodal (upper) and mixed (lower) size distribution disk around Kepler-34. White line shows total number of collisions.}
\label{fig:k34coll}
\end{figure*}

The formation of a static eccentric planetesimal ring in Kepler-16 leads to a similar but suppressed wing effect as compared to the full asymmetric case. Planetesimals orbiting near the ring apoapsis have a reduced orbital velocity and those at periapsis have an increased velocity. Since Kepler-34 has a circular ring at the density peak there is no change to the orbital velocities of the planetesimals and they follow the Keplerian velocity. The lack of departure from the Keplerian orbital velocity and the static nature of the planetesimal rings lead to no significant difference to the gas free case when looking at the impact velocities. It it unsurprising then that the collision type occurrence show no significant change from the gas free case for both Kepler-16 and Kepler-34 as shown in Figures \ref{fig:k16coll} and \ref{fig:k34coll}. 

\subsection{Role of Size Distribution}

A common simplification to many models of planetesimal growth is the assumption of a unimodal size distribution. One reason for removing the complication of a realistic, multi-modal size distribution is the difficulty in understanding what this should be. This is particularly true for circumbinary disks where strong perturbations on the inner disk possibly invalidate commonly used power laws that apply for our unperturbed solar system. Our simplification to this problem, where planetesimals occupy one of ten discrete size bins, is described in Section \ref{sec:method}.

Removing the assumption of same size planetesimals is important in our simulations for three main reasons:

\begin{enumerate}
\item Colliders of equal mass will disrupt more easily due to enhanced momentum coupling. This will lead to a bias in the number of erosive collisions. EDACM uses mass-ratio as an input to collision type determination and thus is susceptible to this bias.\\
\item Colliders with different sizes will lead to larger range of impact velocities since encounter velocities between planetesimals are set by gravitational focusing as well as binary/gas forces.\\ 
\item EDACM uses impact parameter as a quantifier for collision determination. Equal size colliders are more likely to undergo hit-and-run collisions since the critical angle is more likely to be exceeded when the target and projectile have comparable radii. Therefore introducing a more realistic size distribution will help to remove the bias in hit-and-run encounters. 
\end{enumerate}

In both Kepler-16 and Kepler-34 (see Figures \ref{fig:k16coll} $\&$ \ref{fig:k34coll}) we find that introducing this mixed size distribution reduces the number of supercatastrophic collisions and increases the number of erosive events and erosive hit-and-runs. This is consistent for all gas implementation cases. This is expected since planetesimals previously classified as involved in highly erosive (supercatastrophic) events are now less easy to disrupt but still have large values of impact parameter from the orbit crossing events. They therefore become relegated to standard erosion and bounce like hit-and-runs with an erosive element since they erode the smaller projectile in the process.

\section{Discussion}\label{sec:dis}

In this paper we have performed hybrid simulations of circumbinary protoplanetary disks in an attempt to determine the importance of gas disk gravity on the ability for planetesimals to undergo growth-enabling collisions. Our work ties together the pure $N$-body planetesimal dynamics and growth of \cite{lines14} and the hydrodynamical simulations of circumbinary gas disks of \cite{lines15}. The work is comparable to that of \cite{marzari13} who investigate the effects of gas disk gravity on the ability for planetesimal to accumulate in the Kepler-16 system. Our work differs to \cite{marzari13} in a number of ways:

\begin{enumerate}

\item We do not consider the effects of thermal evolution on the gas disks structure and evolution.

\item Our work uses the full implementation of the collision model EDACM \cite{leinhardt12} which allows for the accurate determination of collision outcomes based on multiple factors, and not just impact velocity.

\item We consider 10$^5$ interacting planetesimals with a radius of $R_p \approx$ 270 km, as opposed to 400 $R_p \approx$ 5-25 km test particles. We thus probe an entirely different size regime that considers planetesimals to have somehow grown to this size prior to the start of the simulation.

\end{enumerate}

Our results show that it is imperative to include the full precessing asymmetric gas disk potential when considering the evolution of planetesimal dynamics in circumbinary disks. For both systems we have explored, Kepler-16 and Kepler-34, gravitational effects from the asymmetric gas disk drive up eccentricities with values often exceeding four times that of the background eccentricity set by the dynamical and secular forcing from the binary. The gas surface density, shown in Figure \ref{fig:fluid}, takes a similar shape for both Kepler-16 and Kepler-34; an eccentric ring of material with a high density build up around the inner disk apoapsis and an asymmetric cavity interior to this. Once the planetesimals have reached quasi-steady-state they begin to adopt this shape as those orbiting within the cavity are drawn outwards by the gravitational attraction of the high density gas disk inner edge. Gas gravity not only focuses planetesimals onto eccentric orbits, but particularly around the density peak also efficiently aligns planetesimal pericenters. This means that orbit crossings are far less common in this region since planetesimals are well phased. 

\subsection{Kepler-16}

In Kepler-16 there is an order of magnitude drop in impact velocity at around 1.5 au. This corresponds to a significant drop in the proportion of erosive collision types from this orbital radius, as can be seen in Figure \ref{fig:k16coll}. Particularly problematic however is the region interior to this (1.0 au - 1.4 au) where planetesimals have largely inflated eccentricities but do not undergo strong orbital alignment. Here, eccentric orbits can overlap contributing to large collision velocities during physical interactions. This corresponds to an overwhelming proportion of erosive collisions that reaches  60$\%$ at 1.2 au for the unimodal size case and 70$\%$ for the mixed size distribution.

Erosive collisions either match, or more often, surpass the number of growth enabling (perfect merging and partial accretion) collisions from the inner edge of the disk at 0.6 au to the region of efficient planetesimal pericenter alignment at 1.5 au. Beyond this radius collisions quickly become predominantly accretion based, consistently above 50$\%$ out to the disk edge at 3.0 au. This would suggest that Kepler-16b at 0.7 au could not have formed in-situ, supporting previous hypotheses \citep{lines14,paardekooper12,meschiari12a} that circumbinary planets must have formed further away from the binary barycenter and then migrated to their present location. In this case we find that Kepler-16b couldn't have formed interior to 1.5 au.

There is a noticeable change in planetesimal dynamics and collision outcomes when using only a static axisymmetric gas potential. Planetesimals are pulled onto an eccentric ring at the location of the gas density peak, which despite raising eccentricities across the disk does not lead to a change in collision outcomes over the gas free case. This can likely be attributed to the way planetesimal percenters are aligned over most of the disk, which phases planetesimal orbits such that relative velocities and hence collision velocities are kept low. 

\subsection{Kepler-34}

Since the density peak in the axisymmetric surface density occurs further out than in Kepler-16, it overlays in the planetesimal disk in a region of low number density. This means that the gravitational focusing effect of the gas gravity is weak. For this reason we again find very little modification to the planetesimal velocities and the collision occurrences remain similar to that of the gas free case. For the asymmetric case, similarly because the density peak in Kepler-34 occurs further out than in Kepler-16, it leads to a more eccentric ring. This results in planetesimal being perturbed by the gas over a greater radial extent of the disk and can be seen to increase eccentricities to e $>$ 0.2 between 1.4 au and 2.2 au (Figure \ref{fig:qss}).

Consequently, collisions are dominantly erosive from 1.0 au to 2.0 au with erosion typically accounting for over 50$\%$ of all collisions within this range (Figure \ref{fig:k34coll}). Interior to 1.0 au the low gas surface density has little impact on the collisions and instead the encounter velocities and hence collision occurrences are set by the binary forcing. This is clear from Figure \ref{fig:k34coll} as the high proportion of disruptive collisions from 0.6 au to 0.7 au is also seen in the gas free scenario. In a small window from 0.8 au to 1.0 au erosive collisions are equally matched by growth enabling ones. However the majority of these erosive collisions are supercatastrophic meaning that planetesimals are being ground back down into much smaller bodies and eventually dust. Therefore planetesimals would not only struggle to accrete as much as they lose in this region, but they would likely continue to undergo net erosion. From our results, we suspect that Kepler-34b, which lies at 1.1 au could not have formed in-situ and would have migrated from outside of 2.0 au.

It is important to note that the collision occurrence values in the gas free case are much more favourable to planetesimal growth than those found in \cite{lines14} since the lower resolution of these simulations, and hence increased size of planetesimals, means that planetesimals can grow more easily. This is a simple consequence of larger planetesimals having a larger gravitational binding energy, supporting them against disruption. In this vein it is also important to highlight that our results indicate an extremely hostile disk in the presence of the asymmetric gas gravity despite these large initial planetesimals.

\subsection{Self-Gravity}

The role of inter-planetesimal self-gravity should not be underestimated. We have shown in Figure \ref{fig:grav} that including full gravitational effects between disk bodies appears to change, significantly, the eccentricity distribution. In the axisymmetric gas disk case, the high density ring focuses planetesimals onto a narrow annulus which becomes self-supported by inter-particle gravity. This inter-particle gravity supported ring then has its eccentricity increased by interaction with the binary. It is not surprising that particle gravity plays such an important role in our simulations as the initial size and mass of our planetesimals is large. Full numerical simulations are necessary to correctly account for gravitational focusing effects and gravity-supported structures. It should be noted however, that planetesimal self-gravity may not be the sole factor in the eccentricity inflation. When the mass of the planetesimal disk is reduced, it is likely the stellar binary ceases to precess, and hence it could be the precession of the binary that drives the differences observed in Figure \ref{fig:grav}.

The role of self-gravity was also highlighted in \cite{lines14} where collision outcomes were changed by reducing self-gravity that did not model correctly the dynamics between planetesimal on close approach. Therefore we advise caution for future simulations that do not take self-gravity into account for at least large ($R_{\textrm{p}}$ $>$ 100 km) planetesimals.

\subsection{Gas Drag}

Despite the large size of the planetesimals, they still feel gas drag due to the large eccentricities gained through interaction with gas disk gravity. Using a simple analytical prescription for the implementation of aerodynamic drag from an axisymmetric disk, we perform preliminary simulations with drag enabled in an attempt to quantify the relevance of the drag force in future work. We find that the inclusion of axisymmetric drag has only a marginal impact on the dynamics of collisions between planetesimals, with gravitational effects dominating over drag forces by a few orders of magnitude.

\section{Summary and further work}\label{sec:summary}

Gas feedback on planetesimals is often a missing element from simulations of planetesimal growth in protoplanetary disks, due to the non-trivial nature of unifying these two elements. Our work has succeeded in hybridising 2D hydrodynamical simulations performed with FARGO-ADSG with 3D, high resolution, inter-particle gravity-enabled $N$-body simulations of circumbinary planetesimal disks. Using the state-of-the-art collision model EDACM we have identified regions of disks around Kepler-16 and Kepler-34 where planetesimal growth can occur. We highlight that neither of these systems could have formed in-situ in the presence of a precessing asymmetric disk, even for a disk half the mass of the minimum mass solar nebula. We confine net-growth regions to $r$ $>$ 1.5 au for Kepler-16 and $r$ $>$ 2.0 au for Kepler-34 placing both planets at half this critical radius. Our work corroborates that of \cite{marzari13} who find that the eccentric gas disk gravity inhibits accretion by increasing planetesimal eccentricities and removing their perihelia alignment, and also \cite{meschiari12b} who find that a turbulence gas disk prevents planetesimal accumulation by a similar mechanism.

Several improvements can still be made and will be addressed in future work. These include a) performing further hydrodynamical simulations to solidify our understanding of the structure and evolution of the gas disk by including the heating and cooling of the gas b) increasing the resolution of the $N$-body disk to probe alternate planetesimal size regimes c) improving the handling of collision dust produced during erosive collisions d) performing further investigations into the the importance of self-gravitating planetesimals and e) including drag from a precessing asymmetric disk using a method similar to that used for reconstructing the potential in this paper.

\begin{acknowledgements}
S.L and Z.M.L are supported by the STFC. P.J.C is grateful to NERC Grant NE/K004778/1. SJP is supported by a Royal Society University Research Fellowship. The authors acknowledge the University of Bristol Advanced Computing Research Centre facilities (https://www.acrc.bris.ac.uk/), and the use of Bluecrystal Phase III which was used to carry out this work.
\end{acknowledgements}


\bibliographystyle{mn_new2}
\bibliography{thesis}

\end{document}